\documentclass{aastex}
\usepackage{amssymb}

\begin{document}

\title{The 10 Meter South Pole Telescope}

\author{
J. E. Carlstrom,\altaffilmark{1,2,3,4} P. A. R. Ade,\altaffilmark{5} K. A. Aird,\altaffilmark{6}
B. A. Benson,\altaffilmark{7} L. E. Bleem,\altaffilmark{1,4} S. Busetti,\altaffilmark{8}
C. L. Chang,\altaffilmark{1,4} E. Chauvin,\altaffilmark{8} H.-M. Cho,\altaffilmark{9}
T. M. Crawford,\altaffilmark{1,2} A. T. Crites,\altaffilmark{1,2} M. A. Dobbs,\altaffilmark{10}
N. W. Halverson,\altaffilmark{11} S. Heimsath,\altaffilmark{6}
W. L. Holzapfel,\altaffilmark{7} J. D. Hrubes,\altaffilmark{6}
M. Joy,\altaffilmark{12} R. Keisler,\altaffilmark{1,3}
T. M. Lanting,\altaffilmark{10} A. T. Lee,\altaffilmark{7} E. M. Leitch,\altaffilmark{1,2}
J. Leong,\altaffilmark{13} W. Lu,\altaffilmark{13} M. Lueker,\altaffilmark{7} D.~Luong-Van\altaffilmark{1,2}
J. J. McMahon,\altaffilmark{1,4} J. Mehl,\altaffilmark{7} 
S. S. Meyer,\altaffilmark{1,2,3,4} J. J. Mohr,\altaffilmark{14}
T. E. Montroy,\altaffilmark{13} S. Padin,\altaffilmark{1,2} T. Plagge,\altaffilmark{7}
C. Pryke,\altaffilmark{1,2,4} J. E. Ruhl,\altaffilmark{13} K. K. Schaffer,\altaffilmark{1,4}
D. Schwan,\altaffilmark{7} E. Shirokoff,\altaffilmark{7} H. G. Spieler,\altaffilmark{15}
Z. Staniszewski,\altaffilmark{13} A. A. Stark,\altaffilmark{16} C. Tucker,\altaffilmark{5} K. Vanderlinde,\altaffilmark{1,2}
J. D. Vieira\altaffilmark{1,3} 
and R.~Williamson\altaffilmark{1,2}
}

\altaffiltext{1}{Kavli Institute for Cosmological Physics,
University of Chicago,
5640 South Ellis Avenue, Chicago, IL 60637}
\altaffiltext{2}{Department of Astronomy and Astrophysics,
University of Chicago,
5640 South Ellis Avenue, Chicago, IL 60637}
\altaffiltext{3}{Department of Physics,
University of Chicago,
5640 South Ellis Avenue, Chicago, IL 60637}
\altaffiltext{4}{Enrico Fermi Institute,
University of Chicago,
5640 South Ellis Avenue, Chicago, IL 60637}
\altaffiltext{5}{Department of Physics and Astronomy,
Cardiff University,
CF24 3YB, UK}
\altaffiltext{6}{University of Chicago,
5640 South Ellis Avenue, Chicago, IL 60637}
\altaffiltext{7}{Department of Physics,
University of California,
Berkeley, CA 94720}
\altaffiltext{8}{General Dynamics Satcom Technologies,
2205 Fortune Drive, San Jose CA 95131}
\altaffiltext{9}{National Institute of Standards and Technology,
Boulder, CO 80305}
\altaffiltext{10}{Department of Physics,
McGill University,
3600 Rue University, Montreal, Quebec H3A 2T8, Canada}
\altaffiltext{11}{Department of Astrophysical and Planetary Sciences and Department of Physics,
University of Colorado,
Boulder, CO 80309}
%\altaffiltext{12}{Atacama Large Millimeter Array,
%Avenida El Golf 40, Piso 18, Las Condes, Santiago, Chile}
\altaffiltext{12}{Department of Space Science, VP62,
NASA Marshall Space Flight Center,
Huntsville, AL 35812}
\altaffiltext{13}{Department of Physics,
Case Western Reserve University,
Cleveland, OH 44106}
\altaffiltext{14}{Department of Astronomy and Department of Physics,
University of Illinois,
1002 West Green Street, Urbana, IL 61801}
\altaffiltext{15}{Physics Division,
Lawrence Berkeley Laboratory,
Berkeley, CA 94720}
\altaffiltext{16}{Harvard-Smithsonian Center for Astrophysics,
60 Garden Street, Cambridge, MA 02138}

\email{jc@kicp.uchicago.edu}

\begin{abstract}
The South Pole Telescope (SPT) is a 10 m diameter, wide-field, offset
Gregorian telescope with a 966-pixel, multi-color, millimeter-wave,
bolometer camera. It is located at the Amundsen-Scott South Pole station in
Antarctica. The design of the SPT emphasizes careful control of spillover
and scattering, to minimize noise and false signals due to ground pickup.
% achieve low noise and low offsets for large-area surveys
%of the cosmic microwave background (CMB) radiation. 
The key initial
project is a large-area survey at wavelengths of 3, 2 and 1.3 mm, to 
detect clusters of galaxies via the Sunyaev-Zeldovich effect and to 
measure the small-scale 
%high-$l$ 
angular power spectrum of the cosmic microwave background (CMB). 
The data will be used to characterize the primordial matter power spectrum and to 
place constraints on the equation of state of dark
energy.  A second-generation camera will measure the polarization of the 
CMB, potentially leading to constraints on the neutrino mass and the energy 
scale of inflation.
\end{abstract}

\maketitle

\section{\label{Introduction}Introduction}

Observations of the cosmic microwave background (CMB) radiation have
enormous power to address fundamental questions in cosmology.
Primary temperature and polarization anisotropies in the CMB provide
a unique view of the primordial plasma, while secondary anisotropies yield
information about the structures that have formed in the universe
\citep[e.g.,][]{hu02b}.  The cosmological information encoded in the 
medium- and large-scale CMB temperature anisotropy, measured with 
exquisite precision by the {\it WMAP} instrument
\citep{nolta09}, provides some of the most
important evidence in support of our current cosmological model
\citep{dunkley09}.  This model has been further strengthened by the 
detection and measurement of so-called ``E-mode'' CMB polarization
anisotropy \citep[e.g.,][]{kovac02,page07}.
The next frontiers in CMB anisotropy measurement 
are the small-scale secondary anisotropies and the ``B-mode'' 
polarization signature.  Measuring this next generation of signals is the main
scientific motivation for the South Pole Telescope (SPT).

The most significant source of small-scale secondary CMB anisotropy is 
expected to be the Sunyaev-Zeldovich (SZ) effect, in which CMB photons are scattered
by hot electrons in the deep potential wells of massive clusters of galaxies %(\citealt{sunyaev70}; 
\citep{sunyaev72}.  The SZ effect 
is a potentially powerful tool for finding clusters at high redshift
\citep[e.g.,][]{carlstrom02}.
A sufficiently sensitive SZ cluster survey can produce a large, nearly mass-limited sample of clusters.
Follow-up measurements of photometric redshifts will then allow determination of the
cluster abundance as a function of redshift. This is a sensitive probe
of structure formation, capable of providing strong
constraints on the amplitude of density fluctuations
and on the density and equation of state of dark energy
(\citealt{holder01b}).
%On small angular scales ($l \ga 3000$), CMB temperature anisotropies are expected to be dominated by the
%SZ effect. 
Additionally, a measurement of the angular power spectrum of this
signal will yield constraints on cosmological parameters that are
complementary to those from a cluster survey (\citealt{komatsu02}).

Measurements of the B-mode polarization anisotropy of the CMB probe the
gravitational wave background generated during inflation (\citealt{polnarev85};
\citealt{crittenden93}; \citealt{seljak97};
\citealt{seljak97b}).  This is a unique opportunity to probe physics at the earliest 
epochs and highest energies, but it requires unprecedented raw sensitivity and 
exquisite control of systematics.
The situation is complicated further because gravitational lensing of the E-mode
CMB polarization by
large-scale structure also generates B-mode polarization. Unraveling the
signals will require measurements over a wide range of angular scales
(\citealt{okamoto03}; \citealt{knox02}). The lensing polarization signal is
interesting in its own right. It is another probe of the growth of
large-scale structure, so it can also provide constraints on the equation of
state of dark energy, curvature and the neutrino masses (e.g., \citealt{tegmark05}).

The measurements of the CMB necessary to achieve these science goals involve imaging large areas of sky with high sensitivity at millimeter wavelengths. The best bolometer detectors in this band are already close
to sky-noise limited (\citealt{richards94}; \citealt{holland02}).  Therefore, the SPT aims to improve sensitivity by increasing the number of background-limited detectors, observing at the best site, and minimizing systematic errors. As discussed in \citet{ruhl04}, the key performance features of 
the SPT are:

\noindent (1) $\sim 1$ arcmin beamwidth, to resolve the SZ effect from galaxy clusters. 
This requires an $8\,$m diameter or larger telescope at $\lambda =2\,$mm. The SPT has a
$10\,$m primary with $20\,\mu$m rms surface error to facilitate future submillimeter
observations.  This level of angular resolution will also enable precise measurements 
of the B-mode polarization from gravitational lensing.

\noindent (2) Low scattering, to reduce detector loading and reduce potential systematic errors such as scan-synchronous offsets due to ground pick-up. This leads to an offset optical design with smooth mirrors to reduce scattering and ground shields to control spillover. 
The secondary mirror is cooled to $\sim 10\,$K and surrounded by cooled absorbing baffles
to limit scattered light and loading on the detectors.

\noindent (3) Unprecedented raw sensitivity, so we can quickly
map large areas of sky to useful depths. An SZ survey must cover a large area of the sky
to yield enough clusters to set useful constraints on dark energy---and the amplitude of the 
SZ effect from massive galaxy clusters is typically tens to hundreds of $\mu $K---while a CMB 
polarimeter must measure hundreds of square degrees of sky to $\mu $K
noise levels to enable a measurement of the inflationary B-mode signal.
This sensitivity requirement is met by a telescope with a wide field of view, filled
with background-limited detectors.

\noindent (4) An excellent site for millimeter observing, in order that the survey sensitivity not be limited by atmospheric opacity or sky noise.

%\noindent (4) High sensitivity, because CMB signals are weak. The amplitude of the SZ effect from
%massive galaxy clusters is 
%typically tens to hundreds of $\mu $K;  the amplitudes of the CMB polarization and fine-scale temperature
%anisotropies are much weaker. 

The current SPT camera has a 966-pixel, temperature-sensitive bolometer array 
receiver operating simultaneously in the 3, 2 and
1.3 mm atmospheric windows.  This provides spectral discrimination against
galactic foregrounds and radio and infrared extragalactic sources, and allows us to
separate CMB and SZ signals.  
The camera uses transition edge sensor (TES)
bolometer detectors with a noise equivalent temperature (NET) of $\sim 450\,\mu$K$\sqrt{\rm s}$ (in CMB\ temperature units) at $\lambda =2$ mm. It is located at
the Amundsen-Scott South Pole station, which is one of the best
millimeter and submillimeter wave sites on Earth.
The next camera on the SPT will be a multi-color 
polarimeter with similar, but even greater, raw sensitivity.

In Section~\ref{Site} of this paper, we review the South Pole site.  Sections~\ref{Optics} and \ref{Telescope} describe the design and performance of the optics and telescope, respectively. Section~\ref{Receiver} contains an
overview of the receiver. In Section~\ref{Observations} we describe our observing strategy. Conclusions are presented in Section~\ref{Conclusions}.

\section{\label{Site}Site}

The South Pole is a high, dry site with exceptional atmospheric transparency
and stability at millimeter and submillimeter wavelengths. In winter, the
median precipitable water vapor is $\sim 0.25$ mm (\citealt{chamberlin01}).
% and the zenith opacity at $\lambda =2$ mm is $\sim 0.03$.
The median brightness
fluctuation power at $\lambda =2$ mm is $\sim 31$ mK$^{2}$ rad$^{-5/3}$ in CMB
temperature units  (\citealt{bussmann05}). This is at least
an order of magnitude better than other established terrestrial sites
(\citealt{bussmann05,sayers10}). Temperatures can fall to $-82$ C in winter, which places severe
constraints on the design of exposed components, but weather conditions are
otherwise fairly benign. The average pressure altitude in winter is 3300 m
(cf. 2800 m physical altitude).  Light ($\sim 5$ m s$^{-1}$) katabatic winds
blow from the East Antarctic Plateau most of the time (\citealt{schwerdtfeger84}),
and high winds are rare. The peak recorded wind speed is only $24$ m s$%
^{-1} $. Snow accumulation is $\sim 150$ mm yr$^{-1},$ but local drifting
around surface structures is a problem, so most buildings are elevated. The
ice pack is over 2 km thick and it moves $\sim 10$ m yr$^{-1}.$ 
%It can
%support pressures up to $\sim 30$ kPa without severe tilting or settling.
Transport to the South Pole is via LC130 aircraft, which can carry $11,500$ kg of cargo.
Access is only possible between about 1st November and 14th February of each year.
Air transportation and the short annual construction season are unique and
challenging constraints to fielding instruments at the South Pole.

\section{\label{Optics}Optics}

\subsection{\label{OpticalDesign}Optical design}

\begin{figure}[ht]
\centerline{\scalebox{1}{\includegraphics{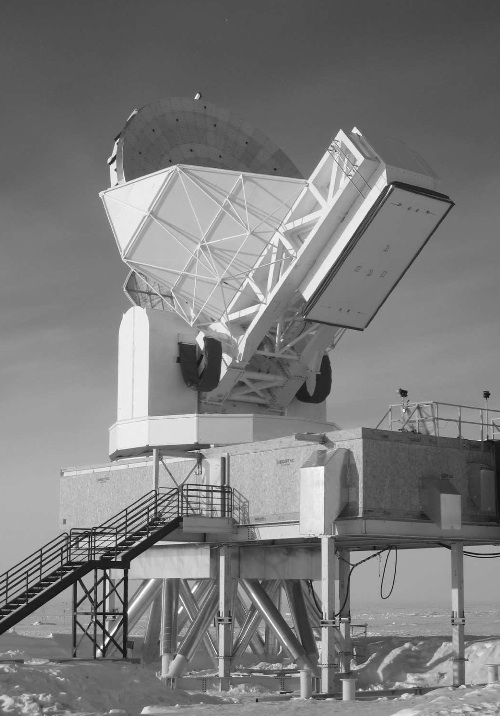}}}
\caption{The 10 meter South Pole Telescope (SPT).}
\label{TelescopePicture}
\end{figure}

\begin{figure}[ht]
\centerline{\scalebox{1}{\includegraphics{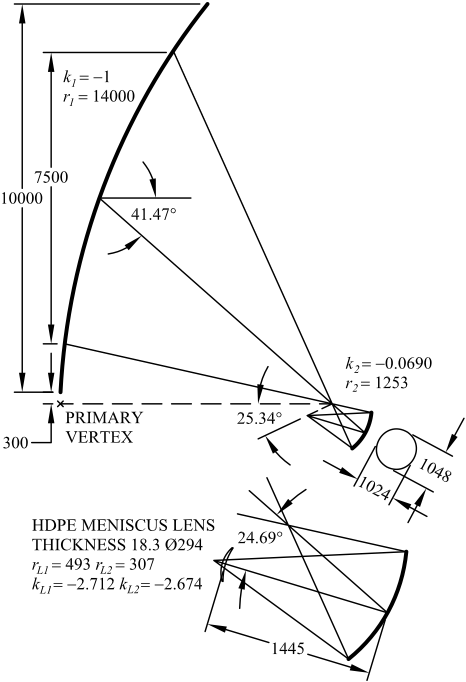}}}
\caption{SPT optics details for (top) the basic Gregorian
telescope with no lens and (bottom) a meniscus lens that makes the final
focus telecentric and gives more uniform illumination of the secondary. For
each surface, $r$ is the radius of curvature and $k$ is the conic constant.
Dimensions are in mm at the operating temperature (ambient for the primary,
10 K for the secondary and 4 K for the lens).}
\label{OpticsLayout}
\end{figure}

The SPT (see Figure~\ref{TelescopePicture}) is an offset classical Gregorian
design. We chose this because: (1) the unblocked aperture minimizes noise and ground
pickup, which is a serious problem for observations of faint, low-contrast
emission such as the CMB; (2) a Gregorian configuration provides an image of
the primary for a chopper or Lyot stop for future receivers;
(3) the secondary in a Gregorian design is concave,
which makes testing of the mirror easier; and (4) the paraboloidal primary
of the classical form allows us to change the focal length of the secondary for
future receivers. An aplanatic design offers a wider field of view
(\citealt{hanany02}) but the focal length of the secondary cannot be
changed much and the focal surface is more curved. The optical
configuration of the SPT is unusually simple because the detectors are at
the Gregory focus (see Figure~\ref{OpticsLayout} and \citet{padin08}).
There are just 2 mirrors
(primary and secondary) and a lens (to make the final focus telecentric and
improve the illumination of the secondary). This scheme gives low loss,
scattering and instrumental polarization, and makes alignment easy. The
field of view is roughly $\lambda \left( \mbox{mm}%
\right) \times 0.7$ degrees and is limited mainly by coma.

The primary has a 10 m diameter aperture with a focal length of 7 m. Prime
focus is 300 mm below the bottom of the primary. This arrangement gives a
reasonable compromise between aberrations, ease of manufacture, and the size
of the secondary support structure.  For more detailed parameters of the telescope
optics, see Figure 2 and Table 1 of \citet{padin08}.

Millimeter-wave telescopes usually have a chopping mirror that quickly scans
or switches the beam to \textquotedblleft freeze\textquotedblright\
atmospheric and gain fluctuations. A chopping secondary is sometimes used,
but telescopes with a wide field of view usually have a flat chopping mirror
at an image of the primary. The image of the primary just after a Gregorian
secondary is a common choice. Wide field designs favor a fast secondary, to
keep the size of the focal plane reasonable, but this gives a poor image of
the primary and increases chop-synchronous offsets. For the SPT, we decided
to abandon a chopper in favor of rapidly scanning the entire telescope.
This works for the low-impedance TES bolometers and frequency domain readout in 
our receiver, but could be a problem for semiconductor bolometers, which are typically
sensitive to vibration from the telescope drives. Avoiding a chopper was
an important choice because it allowed us to make the Gregory focus fast
enough to feed the detectors directly. 

The SPT receiver has wafers of detectors mounted behind a close-packed array
of smooth-wall, conical feedhorns. The spacing between horns is 4.8 mm, which gives reasonable separation between the 4 mm diameter pixels on the detector wafers and provides space for the readout wiring. For optimum coupling to a point source, the horn
aperture diameter should be $2F\lambda $, where $F$ is the final focal ratio 
(\citealt{griffin02}). For $\lambda =2$ mm the $F1.3$  telescope optics are well matched to the horn apertures.
The fast final focus moves the receiver close
to the secondary, which increases the mass of the secondary support
structure. In addition, the small secondary magnification means the mirror
must be tilted through a large angle to compensate aberrations
(\citealt{dragone82}). This requires that the receiver be located close to the beam from the primary.
The receiver-to-beam clearance is a fairly severe constraint in the SPT
design, but it is difficult to avoid without adding mirrors.

To control the illumination pattern of the primary while keeping loading low, the optical system must include a cold stop. The SPT optical design does not have a good image of the primary for a stop, so
we moved the exit pupil to the secondary and surrounded the mirror with cold
absorber. The absorber extends from the secondary to prime focus, and also 
into the receiver, so it functions both as the stop and as a shield around the beam. 
In this scheme, the obvious place for a cryostat window is near prime focus, where
the beam is small. The large cold stop at the secondary does require additional cryogenics, 
and the primary must be a little larger because it is no
longer the entrance pupil. The key advantage is good control of spillover because
the entire beam from prime focus to the detectors is contained inside a
cold, absorbing box.

With the exit pupil at the secondary, there is a trade-off between the size
of the secondary and the size of the primary (\citealt{wilson96}). A smaller
secondary, and hence smaller receiver-to-secondary spacing, results in a
larger shift of the illumination on the primary with field angle, so the
primary must be made larger. This is expensive and it pushes the design
towards a larger secondary. For the SPT, we chose a 1 m secondary, for which
standard machining techniques can achieve $\sim 10$ $\mu $m rms surface
error. The primary must then be 1 m oversize, which fits with our 10 m
aperture and our requirement for a 1 arcmin beam to resolve the SZ effect from
clusters of galaxies at $\lambda =2$ mm.

The focal plane of a Gregorian telescope is curved, so we added a lens just
in front of the detectors to make the final focus telecentric. This lens
reduces the final focal length, allowing us to use a slightly higher
secondary magnification, which increases the clearance between the receiver
and the beam. We optimized the lens to center the illumination
pattern on the secondary for all detectors, which pushes the lens to a meniscus
shape. At $\lambda =2$ mm, the spillover on the stop is $20\pm
0.5\%, $ where the range indicates the variation in spillover across the
detector array. 
%The lens is made of high-density polyethylene, with
%azimuthal triangular grooves machined on both surfaces to reduce reflections and increase
%the transmission.
The lens is made of high density polyethylene (HDPE) and is thermally linked
to the $4$~K cold head inside the receiver cryostat.  The shape of the lens design accounted
for a uniform $1.9\%$ contraction of the HDPE as it cools to its operating
temperature, which was measured in the lab to be $6$-$7$~K.  
The surfaces of the lens are machined with circumferential triangular
grooves to minimize reflections.  These grooves are 
$0.56$~mm wide and $0.64$~mm deep, which is approximately a quarter 
wavelength for our $\lambda=2$~mm detectors.  These groove dimensions 
were optimized so that the calculated transmission was $> 98\%$ across each three of the 
observing bands.

\subsection{\label{SecondaryStop}Secondary and cold stop}

\begin{figure}[ht]
\centerline{\scalebox{0.35}{\includegraphics{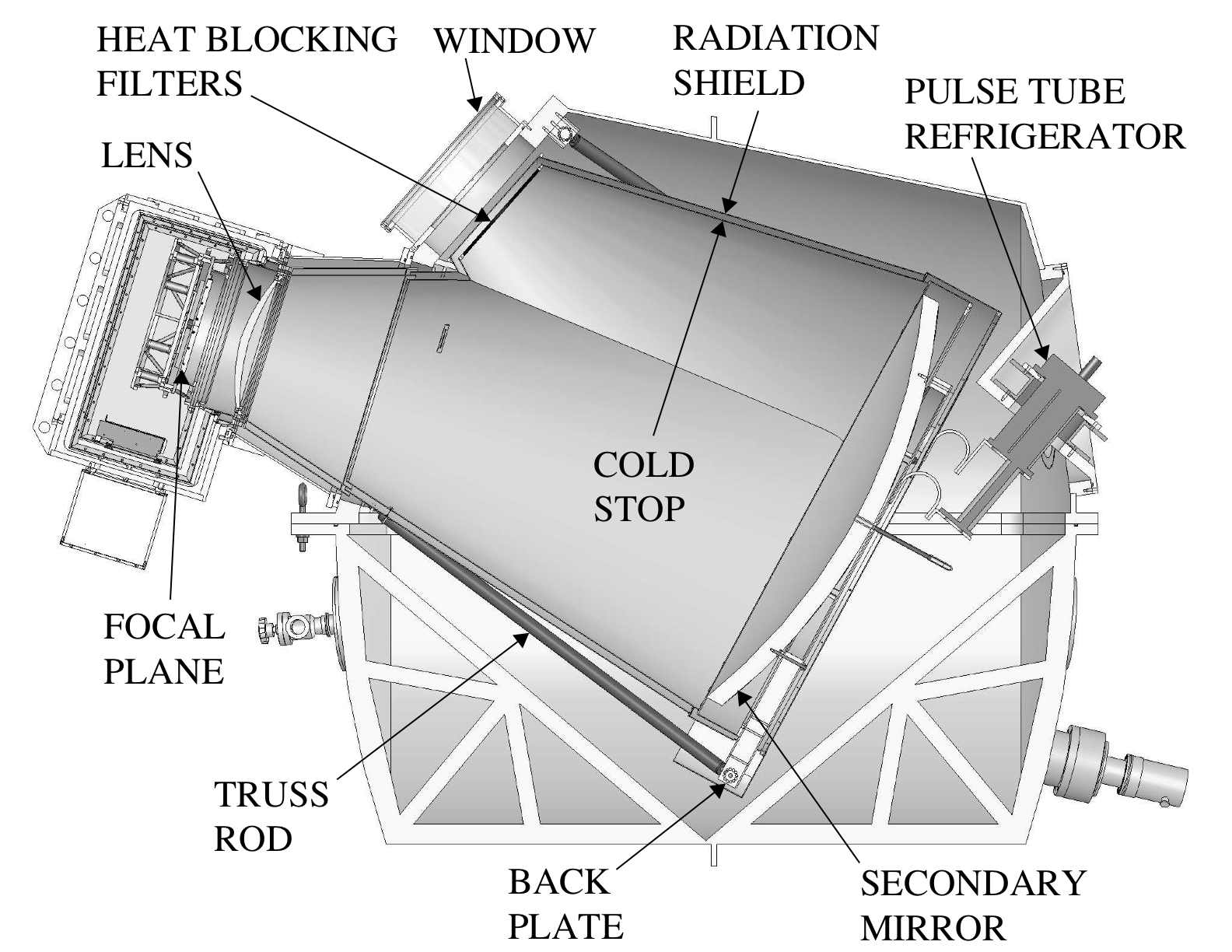}}}
\caption{Section through the receiver and optics cryostats.}
\label{CryostatSection}
\end{figure}

The SPT cryostat has two independent sections that share the same vacuum
space (see Figure~\ref{CryostatSection}). The optics cryostat contains the
secondary mirror, most of the cold stop, and the window and associated
heat-blocking filters. The receiver cryostat contains the lens,
the band-defining filters and the detectors. Each cryostat has its own
refrigerator system. This arrangement allows us to test the receiver without the optics cryostat and to change receivers without disrupting the secondary. The secondary and cold stop are supported by a truss that attaches
to the receiver mounting flange on the optics cryostat. The detectors are
attached to the other side of that flange via a cone and truss. The
separation between the secondary and the detectors is fixed, but the
cryostat assembly is mounted on an optical bench that can be moved $\pm 25$
mm in any direction to maintain alignment with the primary (see Figure~\ref%
{BenchPicture}). This allows us to compensate gravitational flexure of the
secondary support structure and changes in the focal length of the primary
with elevation (which are both a few mm over the full elevation range). The
optical bench actuators have a maximum speed of 25 mm min$^{-1},$ so they can
only follow slow changes.

The secondary is a lightweighted (20 kg) aluminum 7075-T6 mirror, 1 m in
diameter $\times $ 50 mm thick. It is attached at 3 points to a triangular back
plate, which is in turn supported by a truss made of 20 mm diameter $\times $ 1 mm wall
stainless steel tubes. The truss rods have preloaded ball joint ends that
allow some movement during cooling. The secondary surface profile error was
initially 11 $\mu $m rms at room temperature (measured using holography at
89 GHz), but this increased to 50 $\mu $m rms when the mirror was cooled. It
is now 23 $\mu $m rms at room temperature. Stress inside the mirror is
likely responsible for these changes. We did stress relieve the blank by
cooling it to 77 K and then slowly warming to room temperature, 3 times,
before the final cut.  However, the first thermal cycles of the finished mirror
were done in a mount that had no compliance between the mirror and its back
plate. We will probably replace the secondary and improve its mount in the future.

The cold stop is microwave absorber (flexible foam sheet 
HR-10\footnote{Emerson \& Cuming, Billerica MA 01821}) cooled to 10 K.
With 20\% spillover, this contributes just a few K to detector background loading.
The absorber is glued to the inside of a shroud made of annealed, i.e., high
thermal conductivity, aluminum 1100. This is surrounded by a radiation
shield made of the same material. Both the shroud and the shield are covered
with 9 layers of superinsulation (NRC-2-Cryolam\footnote{Metallized Products Inc.,
Winchester MA 01890}) to reduce the radiation load. The secondary
end of the 20 kg stop and shield assembly is attached to the mirror back
plate. The other end is attached to the receiver mounting flange with an
axial flexure. The flexure reduces the torque on the mirror support and allows the
end of the stop and shield to move $\sim 3$ mm on cooling.

Metal-mesh, heat-blocking filters (\citealt{tucker06}) are attached to the
stop shroud and radiation shield just behind the 100 mm thick expanded polypropylene foam 
(Zotefoam PPA30\footnote{Zotefoams PLC, Croydon CR9 3AL UK}) cryostat window.
The loss through the window has been measured to be less than 0.5\% at 2.1 mm.
The stop assembly is cooled by a pulse tube refrigerator
(Cryomech Inc., model PT410\footnote{Cryomech Inc., Syracuse NY 13211})
with a capacity of 10 W at 10 K and 80 W at 70 K. The stop cools to 10 K,
with $<1$ K gradient along its length, and the shield cools to 70 K at the
heat-blocking filters and 60 K at the refrigerator end. Cooling time for the
optics cryostat is 3 days.

Changes in the temperature of the cold stop are coupled to the detectors
through the spillover from the feedhorns and this raises two concerns:

\noindent (1) Systematic signals due to temperature fluctuations from the pulse tube 
refrigerator. This turns out to be a small effect because the
fluctuations are strongly attenuated by the low-pass filter formed by the
thermal resistance of the various heat straps and the heat capacity of the
stop shroud and secondary. A thermometer on the stop shroud shows $<1$ $\mu 
$K rms fluctuations at the $1.4\,$Hz pulse tube frequency. In addition, the pulse
tube frequency is very stable in time and can be notched in the incoming data 
with negligible loss of information. 

\noindent (2) Scan-synchronous offsets or systematic signals noise due to
changes in loading on the cold stop from the cryostat window.
The cold absorber has very low thermal conductivity, so changes in loading 
at the mW level could cause mK amplitude systematic signals in the detectors.
This is addressed in the SPT design through careful filtering, which reduces 
the calculated heat load on the stop to $\sim20 $ mW. 
In addition, the Zotefoam cryostat window is opaque to IR radiation ensuring that 
the power reaching the stop is not significantly modulated by the radiation 
coming through the cryostat window.  
% 20mW is estimate from John Ruhl's calculation

Mechanical stability of the stop is also a concern because any movement of
the beam on the stop, filters, or window could change the spillover and generate 
scan-synchronous offsets. This signal is calculated to be $\sim 10$ mK mm$^{-1}$
in the SPT.
The stiffness of the secondary truss is 0.2 mm/$g$, the peak acceleration
during scanning is $0.05g$ ($4^{\circ }$ s$^{-2}$ 7 m from the azimuth axis)
and the time to accelerate the telescope is $\sim 1$ s, so scan-synchronous
offsets are expected to be $\sim 100$ $\mu $K Hz$^{-1/2}$. Despite the fact that
this is well below the detector noise, we discard data taken when the 
telescope is accelerating.

\section{\label{Telescope}Telescope}

\subsection{\label{Primary}Primary}

The primary has 218 machined aluminum (Al Mg 4.5 Mn) panels mounted on a
composite back up structure (BUS). The BUS is made of 24 identical,
wedge-shaped segments that are essentially deep, stiff boxes with thick
facesheets (see Figure~\ref{MountDiagram}). The segment walls have an aluminum
honeycomb core covered with carbon-fiber-reinforced plastic (CFRP). Invar
inserts are glued into the composite to provide attachment points for
fasteners and panel adjusters. A large, stiff, Invar cone behind the BUS
provides an interface to the steel telescope mount, and an Invar cylinder
running from the center of the BUS to the steel mount adds axial stiffness.

Each aluminum panel is $\sim 0.5$ m$^{2}$ $\times $ 60 mm thick. The mass of 
each panel is reduced to $\sim 7$ kg by machining 20 to 40 pockets with 2 mm
walls and leaving a 3 mm facesheet. The in-plane position of a panel is set by 3 horizontal
adjusters. These have a post with an offset ball that engages a slot in the
back of the panel. Panel position perpendicular to the BUS surface is controlled by 
5 vertical adjusters, one near
each corner of the panel and one near the center. 
%The vertical adjusters
%have a differential screw with a $\sim 100$ mm long $\times $ 8 mm diameter rod that
%acts as a radial flexure. 
The vertical adjusters have a differential screw with a 
$\sim 100$ mm long $\times $ 8 mm diameter rod 
that acts as a flexure. The adjusters were designed to
give just a few $\mu $m rms deformation of the panels due to differential
expansion of the panels and BUS.
A 5-point vertical mount allows some bending to
correct low-order surface profile errors, which are typically $10$ $\mu $m.
After setting all the vertical adjusters, the surface profile error of a
single panel is 5--10 $\mu $m rms. The panels were initially installed on
the BUS based on theodolite measurements of the positions of the panel
corners. This gave a surface profile error of 240 $\mu $m rms.
Adjustments based on photogrammetry measurements at EL$=45^{\circ }$ reduced
the surface error to 40 $\mu $m rms, and a final adjustment based on
holography measurements at EL$=0^{\circ }$ 
%(McMahon et al.\ in prep) 
yielded a 20 $\mu $m rms surface
(see Figure~\ref{PrimarySurface}). 
%Gravitational deflection of
%the primary contributes $11$ $\mu $m rms surface error at the EL extremes
%(after moving the receiver to the new focus position and assuming the
%primary was set at EL$=45^{\circ }$). Thermal deformation is $1$ $\mu $m rms
%for a temperature change of 35 K from the set point. The thermal effect is
%so small because the BUS and its Invar mounting adapters (which all have a
%low coefficient of thermal expansion) are attached to the steel mount with
%radial flexures 
Gravitational deflection of the primary contributes $11$ $\mu $m rms surface
error at the EL extremes (by calculation, after moving the receiver to the
new focus position and assuming the primary was set at EL$=45^{\circ }$). Thermal
deformation of the BUS is $1$ $\mu $m rms (by calculation) for a temperature
change of $35$~K from the set point. The thermal effect is so small because
the BUS and its Invar mounting adapters (which all have a low coefficient
of thermal expansion) are attached to the steel mount with radial
flexures (\citealt{hog75}).

Gaps between panels are nominally 1 mm wide at $-30^{\circ }$ C (the
temperature at which we assembled the primary), increasing to 2 mm at $%
-80^{\circ }$ C. The total gap area is $\sim 1\%$ of the primary, so the
gaps would contribute $\sim 2$ K to the receiver noise and $\sim 1\%$ polarization
errors. Both effects are a concern for CMB observations, so we covered the
gaps with 5 mm wide $\times $ 75 $\mu $m thick BeCu strips. These strips have spring
fingers that engage the panel edges to pull the strip flat against the panel
surface and center it in the gap \citep{padin08b}. The cover strips
contribute 11 $\mu $m rms, which increases the surface error of the primary
from 20 to 23 $\mu $m rms. The panels were etched in NaOH to give a matte
surface that scatters visible wavelengths. This reduces solar heating of the
receiver cabin window to a few $\times 100$ W,
preventing damage to the optics cryostat vacuum window if the telescope
is pointed at the Sun.

Ice on the primary would increase the receiver noise and scan-synchronous
offsets, so the panels must be kept clean. Manual de-icing, e.g., brushing
and scraping, is not practical for a 10 m telescope, so we have an
electrical de-icing system with a 50 W m$^{-2}$ flexible heating pad mounted
behind each panel. Total power dissipation for the heater is 4 kW. This
system runs continuously, raising the temperature of the panels 1--2 K above
ambient, which is enough to discourage ice accumulation without
significantly distorting the panels. Any ice that does form on the surface
during a storm clears in about a day.

\subsection{\label{Mount}Mount}

The SPT has an AZ-EL fork mount
with both axes balanced to minimize deflections. The primary and receiver
cabin are supported by a large, L-shaped frame with the EL axis and
counterweight at the corner (see Figure~\ref{MountDiagram}). The EL bearings
are spherical roller bearings mounted on the
sides of the L frame, with stationary shafts that connect to the fork arms.
The AZ bearing is a 3 m diameter roller bearing supported by a stiff (2 inch thick
wall) cone tower. Space constraints in the offset design push the EL
counterweight close to the EL axis, which makes for a large counterweight. The total mass of the SPT is $\sim 300,000$ kg and 20\% of this
is the counterweight. The structure breaks down into pieces that fit inside
an LC130 aircraft for transportation to the South Pole, but welding on site
is difficult so there are many flanges and bolts. These represent $\sim 20\%$
of the total mass.

The receiver cabin is a 5.5 m long $\times $ 2.1 m wide $\times $ 3.4 m high,
insulated,
shielded room on the end of the L frame. It moves with the telescope, but
can be \textquotedblleft docked\textquotedblright\ to the control room for
work on the receiver. Large doors on the bottom of the cabin can be
positioned above a sliding section of the control room roof. A retractable
boot then seals the bottom of the cabin to the outside of the roof. In the
docked position, the cabin is essentially an extension of the control room,
providing easy, warm access to the receiver and associated electronics. The
docked configuration also provides a high ceiling for cryostat work that
requires overhead lifting. The beam from the primary enters the cabin
through a 12 mm thick expanded polypropylene foam window
(Zotefoam PPA30).
%; Zotefoams PLC, Croydon CR9 3AL UK). 
The window has a metal shutter
covered with microwave absorber that can be used as an ambient load.

Ground shields are mounted along the L frame to reflect scattered radiation from the receiver window and primary to the sky. These shields are made of foam-core aluminum
panels. Exposed edges are rolled with a radius of 3 cm, to reduce
scattering, and gaps between the shields and primary are sealed with
reflecting flashing. The shields have doors near the bottom of the primary
to make snow removal easier. We did not install de-icing heaters on the
shield panels, but we may add these in the future because manual
de-icing is difficult.

The mount sits on a 5 m thick ice pad. This was built in 150 mm layers
that were compacted by driving a bulldozer over each lift to give a density
of 0.55--0.60 g cm$^{-3}$. After construction, the ice was left to harden
for a year. The ice now has a Young's modulus of $\sim 2$ GPa and a compressive
strength of $\sim 0.5$ MPa. The weight of the telescope is distributed by a
large, wood raft. This is a regular hexagon, 6.4 m along each side, made of $%
150\times 300$ mm timbers on 300 mm centers, with 19 mm plywood facesheets. It
reduces the static pressure on the ice to $\sim 28$ kPa. Deflection of the
ice and wood when the telescope accelerates is calculated to be $\sim 1$ arcsec.
A 5.2 m high
steel spaceframe tower supports the telescope on the raft. This tower helps
to distribute the load, but its main function is to raise the telescope
above snow grade to reduce drifting around the structure. The legs of the
tower are insulated to reduce the cold load on the control room and
deformation due to solar heating in summer.

\subsection{\label{Drive}Drive}

Each axis has 4 brushless DC motors with gearboxes. These are mounted in
pairs, with a torque bias between the motors in each pair to eliminate
backlash. The 2 pairs of AZ motors are mounted on opposite sides of the fork
and they engage an external ring gear on the AZ bearing. Each arm of the
fork has a pair of EL motors that drive a sector gear mounted on the L
frame. All the motors have disc brakes that can stop the motor shaft at full
torque. The L frame also has hard stops with shock absorbers. The drive can
operate at reduced speed with just 2 motors per axis in the event of a
failure. Both axes have 26-bit optical encoders
(BEI model LS898\footnote{BEI Sensors \& Systems Company, Maumelle AK 72113}).
The AZ encoder is mounted at the base of the cone tower on a reference frame
attached to the foundation mounting pads. It has a $\sim 2$ m long
shaft/flexure that passes through the helical-spring AZ cable wrap to the
bottom of the fork. There are 2 EL encoders, one in each fork arm, and they
also have a $\sim 2$ m long shaft/flexure that runs through the EL bearing
to the L frame. Performance details for the drive are given in Table~\ref{t1}.
The SPT can scan between 2 positions $1^{\circ }$ apart, settling within 3
arcsec of the requested position in 3 s. During a scan, it can follow the
commanded path within 20 arcsec at accelerations up to $2^{\circ }$ s$^{-2}$
(with better performance at lower accelerations). Figure~\ref{TrackingErrors}
shows recorded telescope position and tracking errors during a typical 
CMB field scan.
The actual path can always
be reconstructed within 1 arcsec using the encoder readings.

A metrology system measures changes in the overall tilt of the mount and
deformations in the fork structure. There are 3 metrology subsystems:

\noindent (1) A biaxial tiltmeter (Applied Geomechanics model 711-2A\footnote{Applied Geomechanics Inc., Santa Cruz CA 95062})
mounted
on the pedestal wall measures changes in the tilt of the AZ axis, which are
mainly due to settling of the foundation. We have a second tiltmeter in the
fork, just above the center of the AZ bearing, to measure repeatable tilts in the
bearing. These are $\sim 5$ arcsec p-p and they are recorded in a look-up
table which can be applied in the on-line pointing model.

\noindent (2) EL encoders in each fork arm measure twisting of the L frame
and forks. The average of the two encoder readings provides the elevation
position feedback to the drive system and the difference is used to calculate
the AZ error due to twisting.

\noindent (3) A pair of linear displacement sensors (BEI model LGDT-5\footnote{BEI Sensors \& Systems Company, Maumelle AK 72113})
mounted on each fork arm measure changes in the height and rotation of the
arms relative to a stiff, CFRP reference frame inside the fork (see Figure~\ref{ForkMetrology}).
Rotations of the fork arms cause AZ and EL offsets and changes in height tilt the EL axis.

\noindent Metrology corrections can be applied in the telescope control system,
based on a 10 s running average of the sensor readings. This ensures
stability and low noise, but it means we cannot correct deflections due to
fast motion of the telescope.

The SPT has an optical star pointing system for measuring axis tilts, encoder
offsets and gravitational flexure. Two 75 mm near-IR refractors are mounted
on the edge of the BUS (at 12 and 3 o'clock) and a 3rd is mounted on the end
of the L frame by the receiver cabin. The pointing telescopes have CCD video
cameras connected to a frame grabber in the telescope control computer. We
can automatically measure pointing errors for about 30 bright stars in half
an hour. The rms pointing error in one of these runs is typically 3 arcsec
(see Figure~\ref{PointingResiduals}), but the best fit model varies by a few
tens of arcsec on timescales of a day. This appears to be due mainly to
complicated thermal deformation of the fork arms, so we have improved the
insulation and have
installed temperature sensors to measure thermal gradients in the fork.  We 
have implemented a radio pointing model which combines the information 
from the temperature sensors, the linear displacement sensors, and 
observations of galactic H II regions.  This model confirms the drifts seen 
in the star camera observations, and with it we are able to correct for these
drifts and reconstruct pointing with $\sim 10$-arcsec rms on hour timescales
(which is the typical separation between H II region measurements).  We have
no evidence of pointing drifts on shorter timescales than this.

The telescope control system is based on software written for the Sunyaev-Zel'dovich Array (SZA). Requested (AZ,EL) positions (including precession, refraction, mount
errors etc.) are computed every 10 ms and sent to the drive. Encoder readings
and status information (motor currents, faults etc.) are recorded every
10 ms. Scanning is implemented by adding offsets from a table to the
requested (AZ,EL). Most observations are managed by scripts containing
high-level control commands (e.g., wait until a particular time and then
start a series of scans on the target source, or drive to stow if the
receiver temperature is above some threshold).

The drive motors, brakes, gearboxes, encoders and limit switches are all
warm to improve reliability, and they are easily accessible from the
control room, so they can be repaired in winter when it is difficult to work
outside. The cone tower and the bottom of the fork are inside the control
room, which has a rotating roof attached to the fork. The fork arms are
enclosed by insulated cabins (riding on top of the rotating roof) with
openings to the control room below. The only cold drive parts are the EL
sector gears and the EL bearings.

\section{\label{Receiver}Receiver}

\subsection{\label{FocalPlane}Focal plane}

The detectors in the SPT receiver (Shirokoff et al. in prep.) are arrays of horn-coupled, 
spider-web
bolometers with TES thermometers (\citealt{gildemeister99}; \citealt{gildemeister00}).
A voltage-biased TES exhibits strong electrothermal feedback, resulting in
good linearity and a responsivity that is independent of bath temperature
and optical loading (\citealt{lee96}; \citealt{lee98}). The TES devices
also have low impedance (typically $1\Omega $), so they should be fairly
insensitive to vibrationally induced currents. In addition, the resonant circuits in
series with each bolometer (see Section 5.2) pass only a narrow band of signals 
at several hundred kHz where there are no mechanical excitations. 

The SPT detectors have an Al-Ti TES, with a
transition temperature of $\sim 0.5$ K, mounted near the center of a
spiderweb absorber. This absorber is a 1 $\mu $m thick, suspended silicon nitride
mesh, 3 mm in diameter suspended by six 0.5 mm legs. It is coated with gold to give a sheet resistance
of a few $100\Omega /\square .$ High electrothermal loop gain leads to a
short electrical time constant that can cause instability (\citealt{irwin98}),
so our detectors
also have a gold pad coupled to the TES to increase its heat capacity and
slow its electrical time constant. We use triangular arrays of 161 close-packed
bolometers (see Figure~\ref{WedgePicture}) fabricated on 100 mm diameter wafers.
The wafers are metallized on the back to provide a backshort at $\sim \lambda/4$. 
%The wafer thickness
%for the $\lambda =3$, 2 and 1.3 mm bands was 230, 450 and 450 $\mu $m for
%2007 observations, and  250, 150 and 105 $\mu $m for 2008 and 2009. The 450 $\mu $m
%wafers are $\sim 3\lambda /4$ thick at $\lambda =2$ mm. We changed to $\lambda /4$
%wafers for 2008 to increase the bandwidth and coupling efficiency of the detectors.
%Thin wafers are difficult to handle during processing, so we bond a carrier wafer
%to the metallized back surface to give a total thickness of $\sim 450$ $\mu $m.
The bands are defined by the low-frequency cut off in a short length of circular waveguide
between each smooth-wall conical horn and its detector, and by low-pass, metal-mesh
filters (\citealt{ade06}) mounted in front of each feedhorn array. The focal plane has 6
triangular bolometer arrays, for a total of 966 detectors; however, 
due to readout limitations at most 840 detectors can be active. 

For the 2009 season, one $\lambda =3$ mm array, four 2 mm arrays and 
two 1.3 mm arrays have been installed. Typically, $\sim 700$ detectors pass our performance 
cuts with high sensitivity and noise close to the background limit. 
The typical optical coupling from the sky to the TES is $\sim 25\%$.
The mean noise equivalent CMB temperatures (NET) and yields for each of the three bands in the 
2009 SZ receiver configuration are summarized in Table~\ref{receiver}.

The NETs quoted in Table~\ref{receiver} are evaluated at $\sim 3$~Hz 
(or $\ell \simeq 2000$~to~$4000$ for typical scan speeds).  At lower frequencies, 
single-detector noise levels rise due to atmosphere and, to a lesser degree, 
detector and readout $1/f$ noise.  Figure~\ref{NoisePsds} shows NET as a 
function of (temporal) frequency for typical detectors at each observing wavelength
under typical observing conditions.  NET is shown for data taken while scanning
and while stationary.  The slight increase in low-frequency noise while scanning 
is due to the faster imaging of large-angle atmospheric fluctuations.  

For a typical detector, 
the optical time constant is $\sim 15$ ms.  For typical SPT scan speeds 
of $0.25$~to~$0.5$~degrees/s, this implies a $3$~dB spatial frequency of 
$\ell \simeq 7500$~to~$15000$.  For comparison, the $3$~dB point of a $1$~arcmin 
beam is $\ell \simeq 6700$; i.e., the detector time constants are not limiting our high-$\ell$
response at typical scan speeds.  Furthermore, in the regime in which photon
shot noise is the dominant contribution to the detector noise budget, the noise
is rolled off with the optical time constant just as the signal is, and the signal-to-noise
ratio does not decrease (\citealt{lee98}).  Most of the SPT detectors operate in or near this regime, so
much faster scanning is feasible from a detector noise standpoint.

%For 2008 observations
%we had one $\lambda =3$ mm array, three 2 mm arrays and two 1.3 mm arrays.
%Typically, $\sim 570$ detectors were optically active with reasonable noise
%($\sim 30$ detectors at $\lambda =3$ mm, $\sim 345$ at 2 mm and $\sim 195$ at 1.3 mm; the low yield at $\lambda =3$mm is due to a fabrication error that has been corrected for the 2009 season).
%The average NET is $\sim 0.45$ and $\sim 0.9$ mK s$^{1/2} $ (in CMB temperature units)
%at $\lambda=2 $ and 1.3 mm, respectively, which is very close to the background limit 
%for this configuration.

\subsection{\label{Cryogenics}Cryogenics}

The SPT receiver (Benson et al., in prep) is cooled by a pulse tube refrigerator (model PT415\footnote{Cryomech Inc., Syracuse NY 13211})
with capacity of 1.5 W at 4.2 K and 40 W at 45 K,
and a 3-stage $^{4}$He$^{3}$He$^{3}$He sorption 
refrigerator (model CRC10\footnote{Chase Research Cryogenics Ltd., Sheffield S10 5DL UK})
with a capacity of 80 $\mu $W at 380 mK and 4 $\mu $W at 250 mK.
The sub-K refrigerator is cycled automatically in $\sim 3$ hrs and the hold time is $\sim 36$ hrs.

\subsection{\label{Readout}Readout}

The SPT receiver readout (Dobbs et al. in prep) is a frequency multiplexed SQUID readout with 8 detectors per
SQUID (\citealt{spieler02,lanting05}).  Each TES is
biased with a constant voltage amplitude sine wave, in the 0.3--1 MHz range,
and has a series LC filter, mounted near the focal plane at 250 mK, to
select the appropriate bias frequency from a comb of 8 (see Figure~\ref{ReadoutDiagram}).
In this scheme, only
2 superconducting NbTi wires are needed to connect 8 detectors in the focal
plane to their SQUID, which is mounted at $4$~K. The LC filters all have 16~$%
\mu $H chip inductors, so the filter Q increases with frequency and the
bandwidth ($\sim 5$~kHz in the SPT) is constant. The filter frequency is set
by the capacitor, which is a standard ceramic chip device. The SQUIDs are
100-element series arrays (\citealt{huber01}) with a small input coil to reduce
pickup of spurious signals. These devices have $120$~MHz bandwidth, $500$~V/A
transimpedance and the noise (referred to the input coil) is $2.5$~pA~Hz$%
^{-1/2}$ (cf. $\sim 15$~pA~Hz$^{-1/2}$ bolometer noise). SQUIDs are
extremely sensitive to magnetic fields, so the SQUID boards (each with 8 SQUID arrrays) 
are enclosed in a Cryoperm shield to attenuate any external fields. 
Inside the cryoperm shield, each SQUID is mounted on a Nb film (type 2 superconductor)
to pin any residual magnetic flux. 
To maintain constant voltage bias across the
TES, the input impedance of the SQUID must be small compared with the TES
resistance. This requires that the SQUID be operated with shunt feedback
from the output of the room-temperature amplifier that follows. Negative
feedback also linearizes the SQUID response, reducing intermodulation
between the bias signals. The feedback amplifier has a high gain $\times $ bandwidth
product, so connections between the SQUID and the room temperature
electronics must be short. This is a severe constraint on the mechanical and
thermal design of the receiver.

Bias signals for the detectors are generated by direct digital synthesizers
with low sideband noise. The SQUID amplifier outputs are demodulated by
square-wave mixers, which have low distortion because they contain no
non-linear components. The demodulated signals are filtered by 8-pole,
low-pass, anti-aliasing filters and then digitized with 14 bits resolution.
All the bias synthesizers and demodulator electronics are on 9U VME cards
with 16 channels per card. Each card has a field programmable gate array
that packages the data with a time tag and provides control and monitoring
of the readout configuration. The electronics for a 966-pixel receiver
occupy 3 VME crates (see Figure~\ref{ReadoutPicture}).

\section{\label{Observations}Observations}

Most SPT observations involve scanning the telescope to fully sample the sky
with all the detectors (and hence all of the frequency bands). We generally
scan back and forth in AZ at constant speed, turning around as quickly as
possible, with a step in EL at the end of each AZ scan. The speed in the
linear part of the AZ scan involves a trade off between noise and observing
efficiency. Higher scan speeds move the sky signals to higher frequency in
the detector timestreams, so low-frequency noise from the atmosphere and
receiver become less important, but the observing efficiency is reduced
because a larger fraction of the time is spent turning around. 
%This trade off tends to favor a higher scan speed for a larger field. 
Typical scan speeds for large CMB fields have ranged from 
$0.25$~to~$0.5$~degrees/s.\footnote{Scan speeds are quoted here in 
terms of geodesic distance; speeds in units of degrees of azimuth per second
will be higher by a factor of one over the cosine of the observing elevation.}
The EL steps at the end of each scan are profiled to minimize the excitation of elevation oscillations, which would lead to scan synchronous modulations of the atmosphere. 

A typical observation starts with a small ($\sim 2$~degree) scan in EL 
(to inject a signal from the atmosphere), followed by
observations of a chopped thermal source and H II regions.  In Figure~\ref{rcw38_image} we show the 
SPT images of one of the galactic HII regions used, RCW38.
This set of calibration observations allows us to measure the
detector gains and monitor the opacity of the atmosphere.
The chopped thermal source is a $\sim 1000$ K black-body,
with a 4--100 Hz chopper wheel, connected to a 9.5 mm diameter light
pipe that runs into the optics cryostat and through a hole in the center of
the secondary. We check the pointing by mapping two bright H II
regions near the target field, which takes $\sim 20$ min total, and then we scan
on the target field for a few hours.  The cycle is repeated until the mK
refrigerator in the receiver warms up. Detector outputs and telescope
positions are recorded at 100 Hz, but most of the monitoring (e.g., cryostat
temperatures, weather, receiver readout configuration and optical bench
position) is at 1 Hz. The data rate is typically 30 GBytes/day and all the
data are returned to the US via NASA's Tracking and Data Relay Satellite
System.

About half of our 1st year (2007) observations were performance tests (e.g.,
measurements of beams, focus, pointing, ground pickup and detector noise and
time constants). The rest of the year was spent on a deep observation of a $\sim 7$
deg$^{2}$ field and a $\sim 50$ deg$^{2}$ field to search for galaxy clusters via the SZ effect and
to measure the high-$l$ angular power spectrum of the CMB.  The second year observations concentrated primarily on 
SZ and fine-scale CMB surveys in two $\sim 100$ deg$^{2}$ fields. Targeted SZ observations
toward known galaxy clusters were also pursued.  Since the beginning of the third observing
season, we have spent the vast majority of time on the survey and are currently mapping 
the sky to survey depth at a speed of over $700$ deg$^{2}$~yr$^{-1}$.  
%Now in the third year, we are continuing the SZ survey.  
We reported the first clusters detected via their SZ signature in \citet{staniszewski09} and 
have since published many significant results, including the first SZ-selected 
cluster catalog and cosmological constraints from that catalog, the first detection of 
the power spectrum of secondary CMB anisotropy, a point-source catalog that includes
an exciting new population of high-redshift sources, and observations of the SZ effect
in known galaxy clusters out to unprecedented clustercentric radii 
\citep{vanderlinde10,lueker10,vieira10,plagge10}.

\section{\label{Conclusions}Conclusions}

The South Pole Telescope (SPT), a 10 m diameter, wide-field, offset Gregorian telescope equipped with a 966-pixel millimeter-wave bolometer camera, has
been successfully deployed at the Amundsen-Scott South Pole station.  
The SPT has been operating
at the South Pole since February 2007. It has a simple,
low-loss optical system, with good control of spillover and scattered
radiation. Low noise, small offsets and high throughput, combined with the
exceptional transparency and stability of the South Pole atmosphere, makes
the instrument ideal for survey work and observations of low surface
brightness emission, such as the CMB. The telescope has a 20 $\mu $m rms
surface, and pointing errors of a few arcsec, so it can be used at
submillimeter wavelengths.

The first key project of the SPT is a large scale survey of the sky (1000's of sq.\ degrees) 
at  3, 2 and 1.3~mm to search for the galaxy clusters via the SZ effect and 
to measure the fine-scale anisotropy of the CMB.  
The SZ receiver is now operating with background limited performance in all three 
observing bands.
%Early survey results include the first clusters to be detected via a SZ 
%survey (\citealt{staniszewski09}).

This paper offers a snapshot of the current status of the SPT.  Throughout the life of the telescope it is expected
that several receivers will be deployed for conducting major large-scale survey projects at millimeter through submillimeter wavelengths.
We are currently building a polarimeter
for measuring the B-mode polarization anisotropy of the CMB. 

\bigskip
\bigskip
\acknowledgements We thank R. Hills for many useful discussions on the optical design. We gratefully acknowledge the
contributions to the construction of the telescope from
T.\ Hughes, P.\ Huntley, B.\ Johnson and E.\ Nichols and his team of iron workers. We also thank the 
National Science Foundation (NSF) Office of Polar Programs, the United States Antarctic Program and the Raytheon Polar Services Company for their support of the project.  We are grateful for professional support from the staff of the South Pole station The SPT is supported by the National Science Foundation through grants ANT-0638937 and ANT-0130612.  Partial support is also provided by the  
NSF Physics Frontier Center grant PHY-0114422 to the Kavli Institute of Cosmological Physics at the University of Chicago, the Kavli Foundation and the Gordon and Betty Moore Foundation. 
The McGill group acknowledges funding from the National Sciences and
Engineering Research Council of Canada, the Quebec Fonds de recherche
sur la nature et les technologies, and the Canadian Institute for
Advanced Research. The following individuals acknowledge additional support: J.\ Carlstrom from
the James S.\ McDonnell Foundation, 
K.\ Schaffer from a KICP Fellowship,  J.\ McMahon from a Fermi Fellowship,  Z.\ Staniszewski from a GAAN Fellowship, and A.T.\ Lee from the Miller Institute for Basic Research in Science, University of California Berkeley.

\newpage 
\begin{table}[tbp]
\caption{Drive performance.}
\label{t1}%
\begin{tabular}{llll}
Parameter & AZ & EL & Units \\ \hline
Range & $\pm 270$ & 0--140 & deg \\ 
Maximum speed & 4 & 2 & deg s$^{-1}$ \\ 
Maximum acceleration & 4 & 4 & deg s$^{-2}$ \\ 
Drive gear ratio & 9.319 & 13.410 &  \\ 
Motor gearbox ratio & 579 & 894 &  \\ 
Motor speed & 3600 & 4000 & rpm \\ 
Motor torque & 33 & 33 & ft lb \\ 
Motor inertia & 0.016 & 0.016 & ft lb s$^{2}$ \\ 
Structure inertia & $1.69\times 10^{6}$ & $2.15\times 10^{6}$ & ft lb s$^{2}$ \\ 
Structure inertia at motor shaft & 0.058 & 0.015 & ft lb s$^{2}$ \\ 
Resonant Frequency & 3.7 & 2.5 & Hz%
\end{tabular}
\end{table}

\begin{table}[tbp]
\caption{2009 SZ Receiver Performance}
\label{receiver}
\begin{tabular}{lcc}
Band (mm) & Typical \# Active Bolometers & Typical NET ($\mu\,{\rm K_{CMB}}\sqrt{s}$)  \\ \hline
3 &  120  & 600 \\
2 &  465  & 450 \\
1.3 & 100 & 1200 \\
\end{tabular}
\end{table}

\newpage 

\newpage 

\newpage 

\newpage 
\begin{figure}[ht]
\centerline{\scalebox{1}{\includegraphics{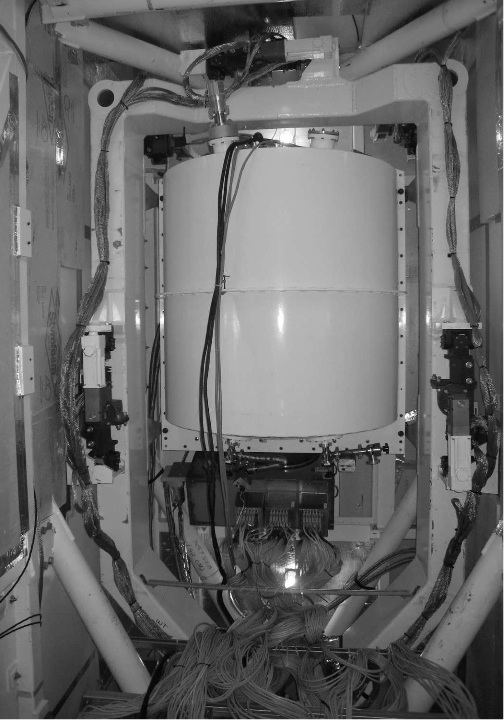}}}
\caption{Optical bench assembly. This view is looking up
into the receiver cabin from the control room. The sky end of the cabin is
at the top. The optics cryostat is the large, white, cylindrical vessel in
the center of the picture and the optical bench is the white frame around
it. The receiver (with many wires running down to the readout electronics)
is just below center. Three of the 6 optical bench actuators are visible at
the top, left and right of the picture. The other 3 actuators connect the
corners of the bench to the cabin roof.}
\label{BenchPicture}
\end{figure}

\newpage 
\begin{figure}[ht]
\centerline{\scalebox{0.45}{\includegraphics{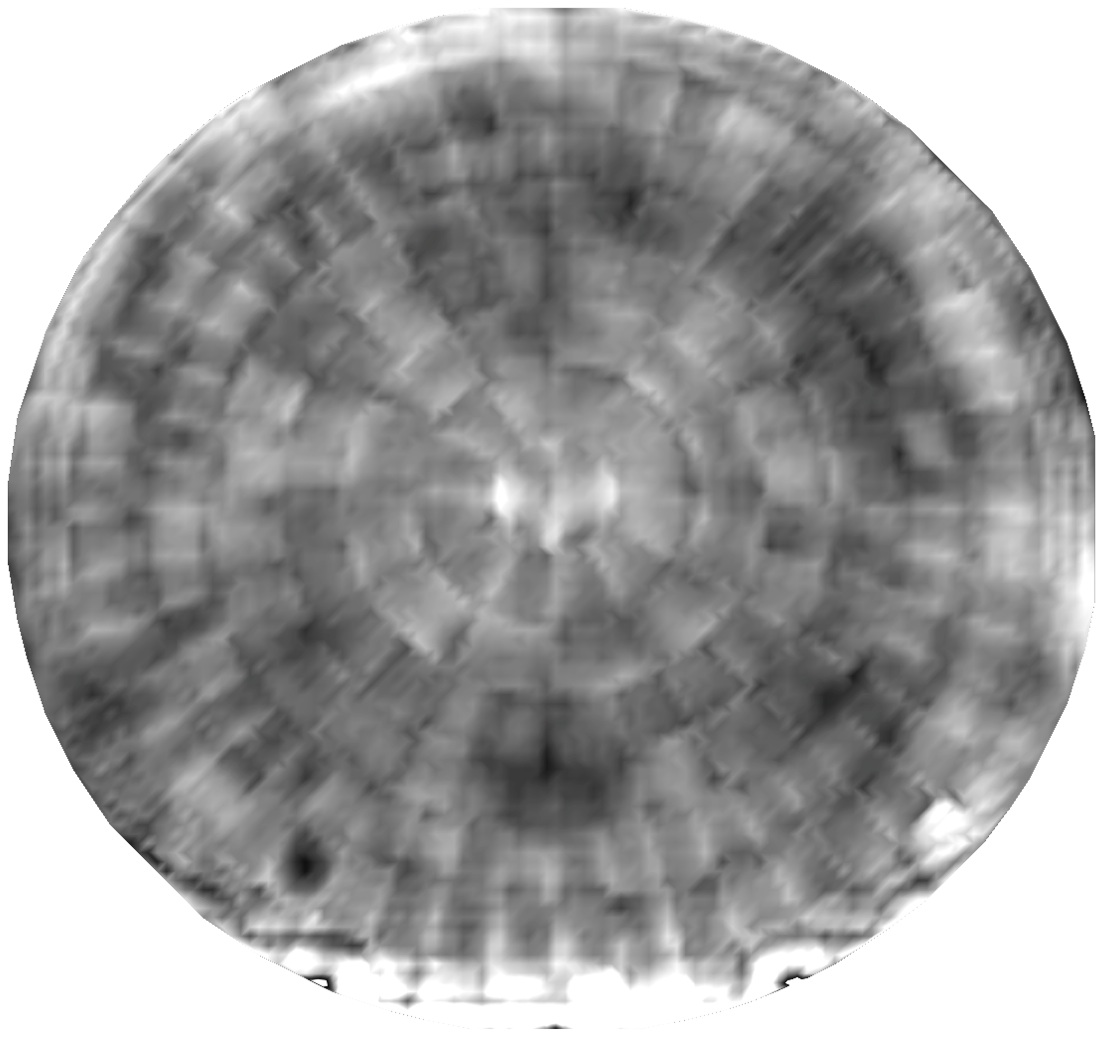}}}
\caption{Primary surface profile errors. The greyscale range is
$+75$ $\mu $m (white) to $-75$ $\mu $m (black). The features at
the bottom are artifacts caused by reflections from the L frame where the
beam enters the receiver cabin near prime focus. Light (i.e., high) stripes along the panel
edges are due to the gap covers. The surface error (with uniform weighting
over the entire primary) is 20 $\mu $m rms.}
\label{PrimarySurface}
\end{figure}

\newpage 
\begin{figure}[ht]
\centerline{\scalebox{0.45}{\includegraphics{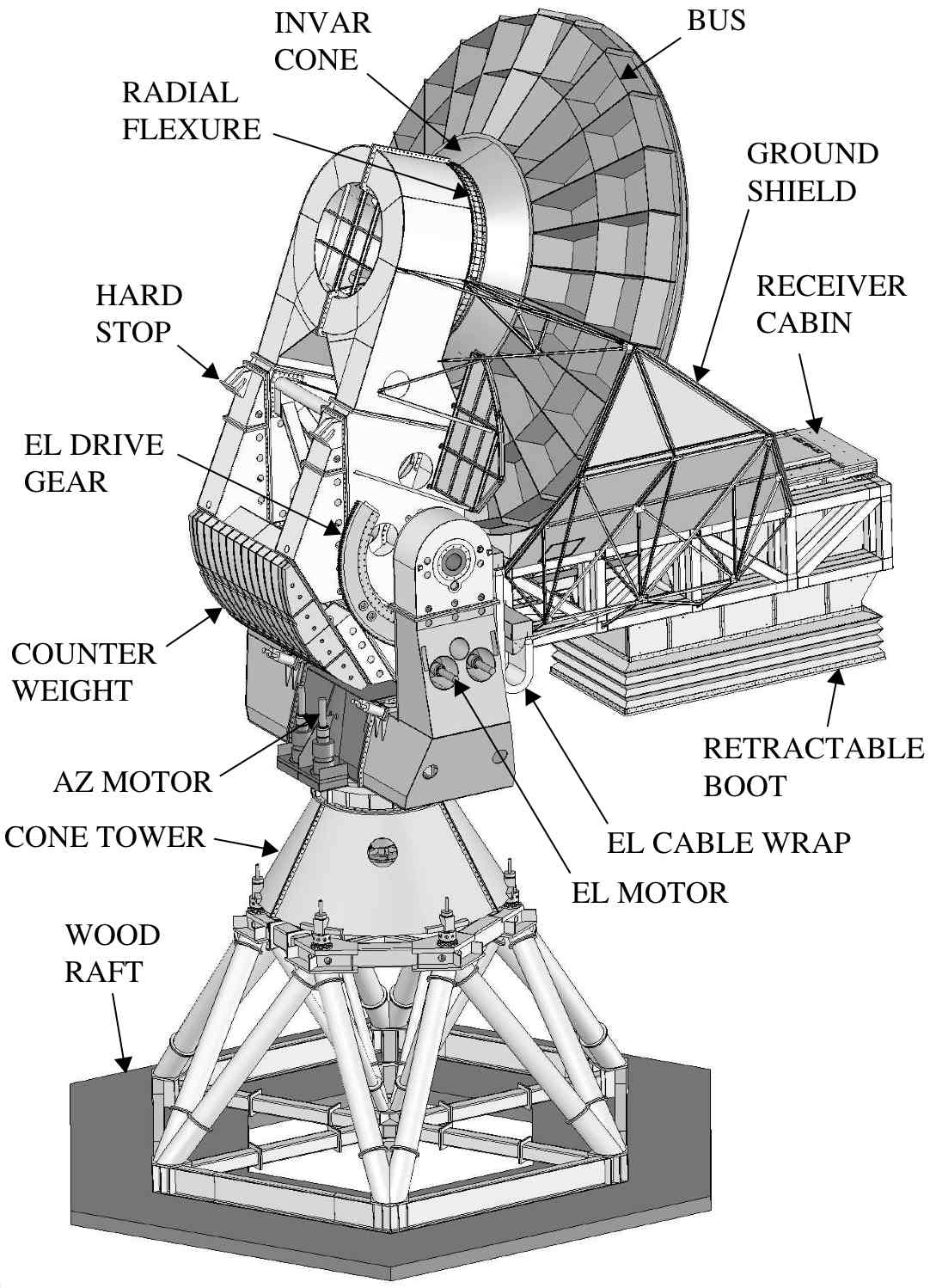}}}
\caption{Mount details. The receiver cabin, fork cabins,
control room rotating roof and the covers on the back of the BUS have been
removed to show the telescope structure.}
\label{MountDiagram}
\end{figure}

\newpage 
\begin{figure}[ht]
\centerline{\scalebox{0.45}{\includegraphics{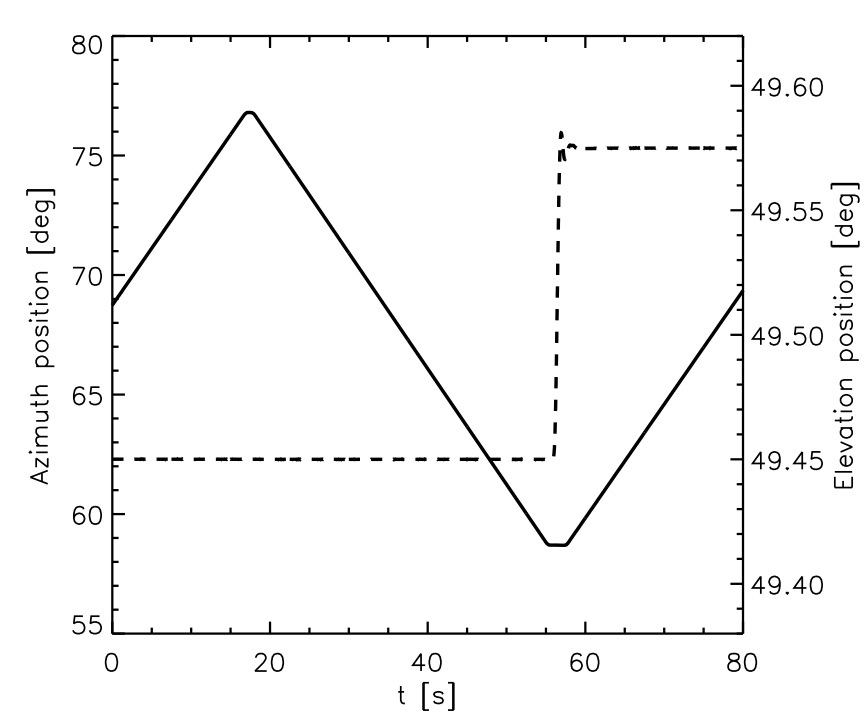}
\includegraphics{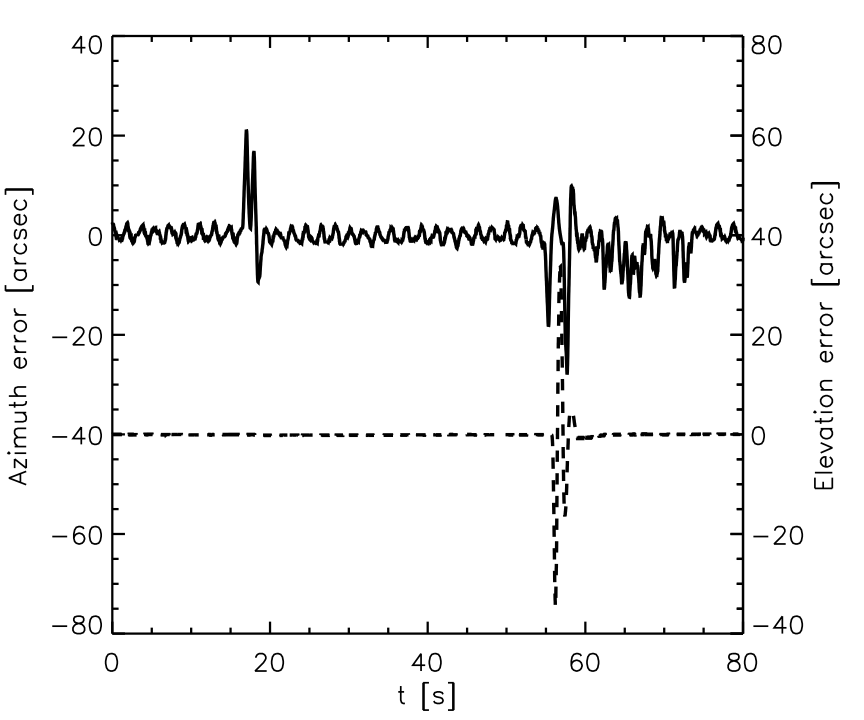}}}
\caption{Recorded telescope position and tracking errors (recorded position minus
commanded position) during a typical CMB field scan.  In both plots, the solid line is 
azimuth, and the dashed line is elevation.  (Note the different scales 
for azimuth and elevation position and the offset between the azimuth and elevation 
error axes).  Tracking errors briefly exceed 20~arcsec during azimuth turnarounds 
and elevation steps, but they quickly settle to well below the 20-arcsec specification 
and remain so throughout the constant-velocity portion of the scan.}
\label{TrackingErrors}
\end{figure}

\newpage 
\begin{figure}[ht]
\centerline{\scalebox{1}{\includegraphics{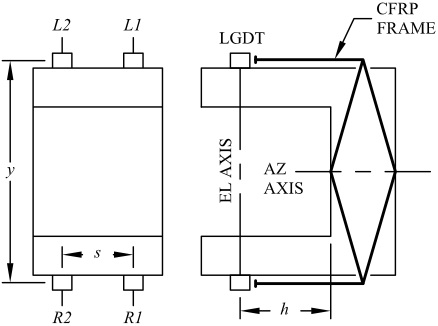}}}
\caption{Fork metrology system. LGDT is a linear gap displacement transducer.
$L1,$ $L2,$ $R1$ and $R2$ are distances (measured parallel to the AZ axis)
from points on the fork to the CFRP frame. If $L1$ and $R1$ are on the receiver
cabin side of the fork, the corrections for AZ and EL offsets and EL axis tilt
are $\Delta AZ=\left(  L1-L2-R1+R2\right)  h/ys  ,$ $\Delta
EL=\left(  L2-L1+R2-R1\right)  /2s $ and $\Delta ET=\left(  L1+L2-R1-R2\right)
/2y $ radians.}
\label{ForkMetrology}
\end{figure}

\newpage 
\begin{figure}[ht]
\centerline{\scalebox{0.75}{\includegraphics{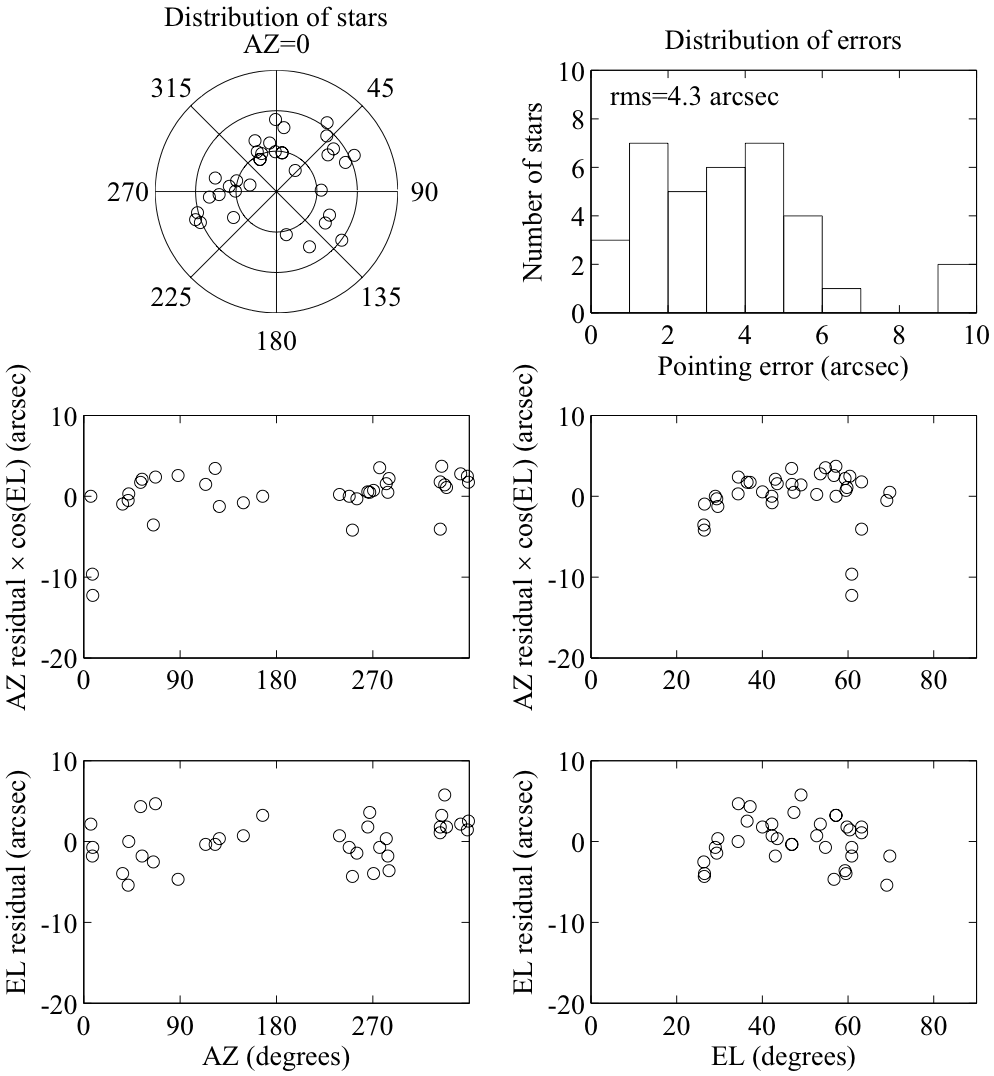}}}
\caption{Pointing errors measured with the optical
telescope on the end of the L frame. This observation of 35 stars was made
on 17-Jun-2007 at 23:08 UT. The plots show AZ and EL residuals after fitting
for AZ and EL axis tilts, AZ and EL encoder offsets, sin(EL) and cos(EL)
flexure and optical telescope cross-EL collimation. The rms of the residuals
is 4.3 arcsec, but this includes $\sim 3$ arcsec rms seeing, so the mount
errors are $\sim 3$ arcsc rms. The 2 outliers near AZ$=0$ are in the Alpha
Centauri system. They consistently have high AZ residuals, possibly because of
an error in the proper motion correction.}
\label{PointingResiduals}
\end{figure}

\newpage 
\begin{figure}[ht]
\centerline{\scalebox{1}{\includegraphics{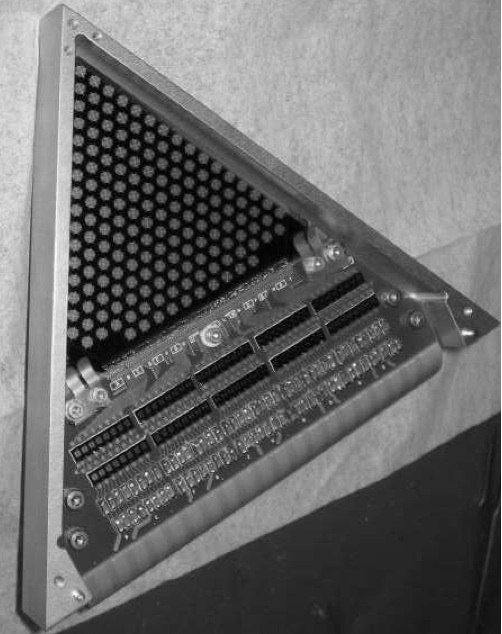}}}
\caption{161 element bolometer array (without feedhorns and
filters). The spacing between bolometers is $\sim 5$ mm. Bias filters for the
frequency multiplexed readout are on the circuit board at the bottom of the
picture.}
\label{WedgePicture}
\end{figure}

\newpage 
\begin{figure}[ht]
\centerline{\scalebox{0.4}{
\includegraphics{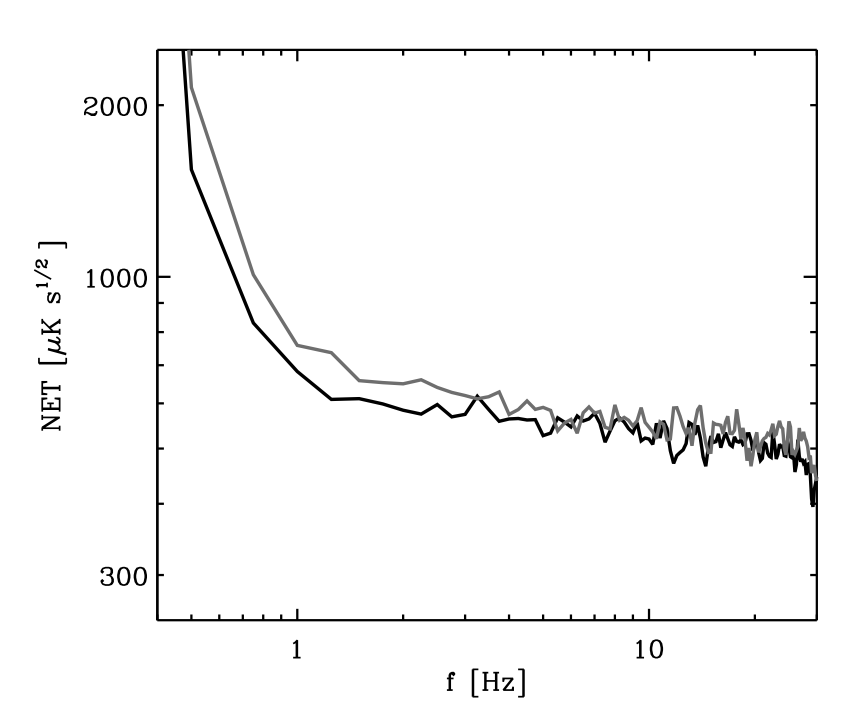}
\includegraphics{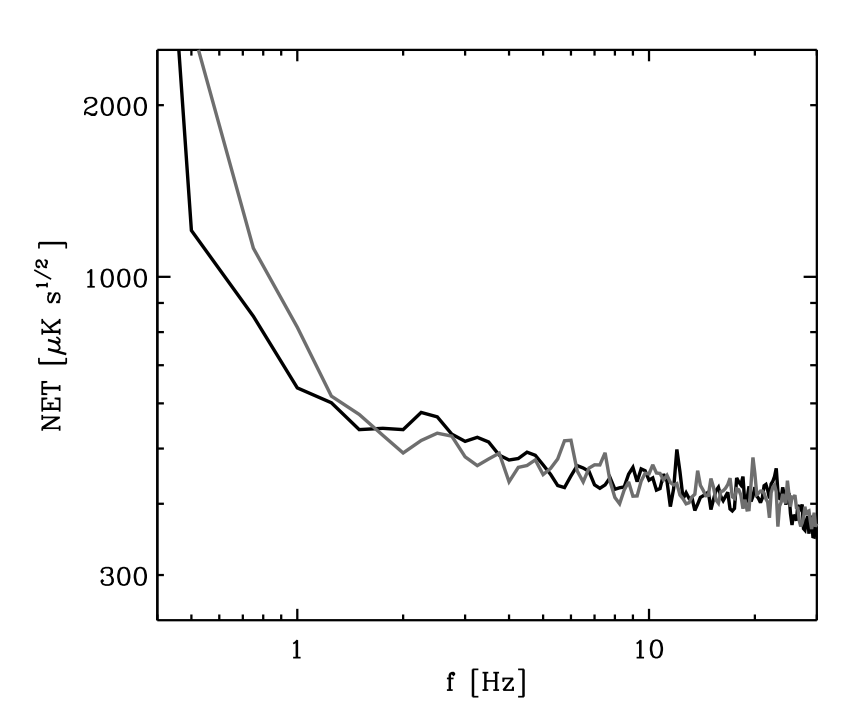}
\includegraphics{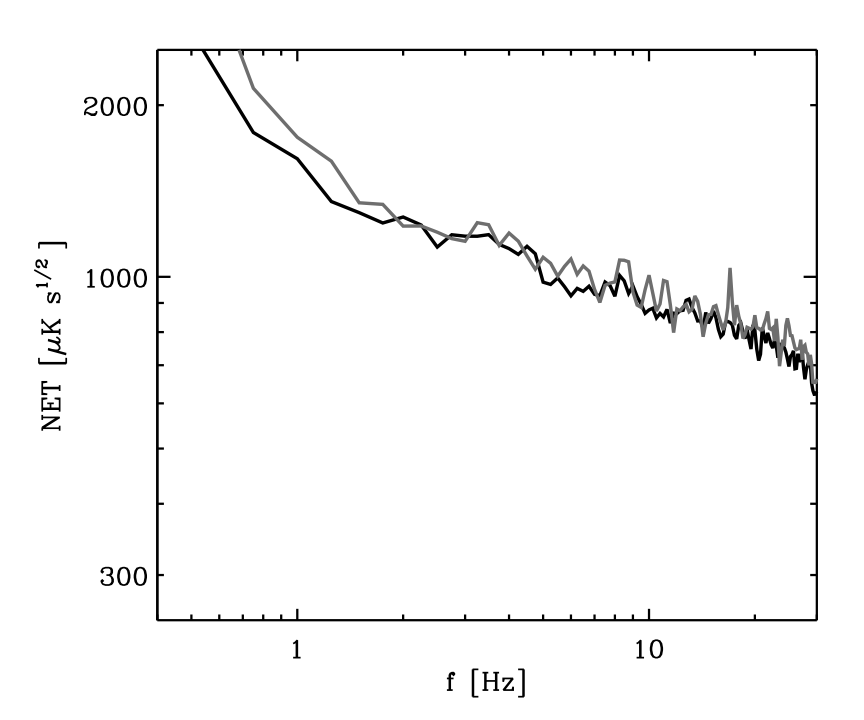}
}}
\caption{NET as a function of frequency for a typical bolometer at 
$\lambda = 3$~mm ({\it left panel}), $2$~mm ({\it center panel}), 
and $1.3$~mm ({\it right panel}).  In each plot, the black trace shows
data taken while the telescope was stationary, while the gray trace 
shows data taken while the telescope was scanning at 
$\sim 0.25$~degrees/s.}
\label{NoisePsds}
\end{figure}

\newpage 
\begin{figure}[ht]
\centerline{\scalebox{1}{\includegraphics{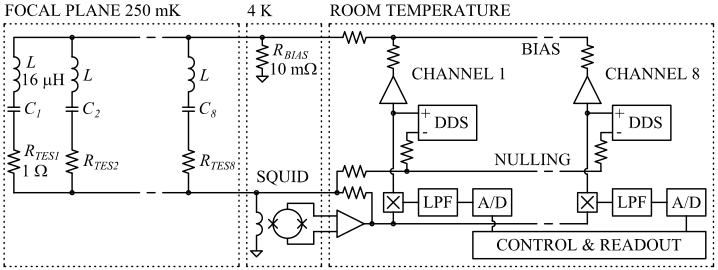}}}
\caption{Frequency multiplexed SQUID readout. DDS is a
direct digital synthesizer, x is a square-wave mixer, LPF is a low-pass,
anti-aliasing filter and A/D is a digitizer. Nulling suppresses the bias
carriers to reduce the dynamic range at the SQUID input.}
\label{ReadoutDiagram}
\end{figure}

\newpage 
\begin{figure}[ht]
\centerline{\scalebox{1}{\includegraphics{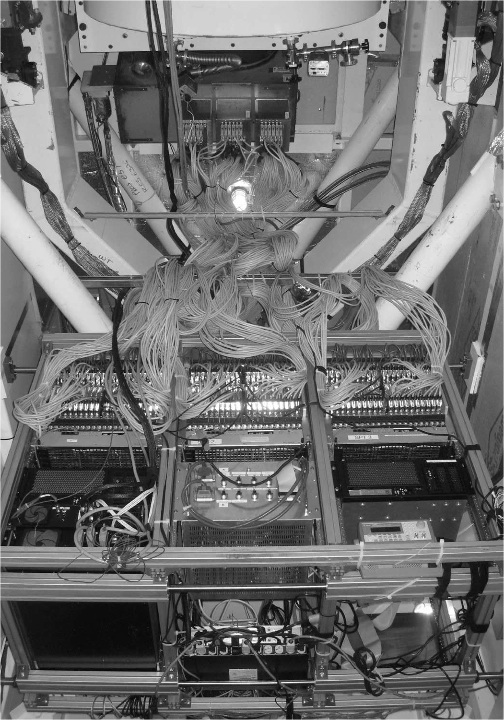}}}
\caption{Readout electronics. The receiver and optics
cryostats are at the top of the picture and the 3 VME crates of bias and
demodulator electronics are at the top of the rack. The 2 white tubes
immediately above the rack are part of the L frame.}
\label{ReadoutPicture}
\end{figure}

\newpage 
\begin{figure}[ht]
\centerline{\scalebox{1}{\includegraphics{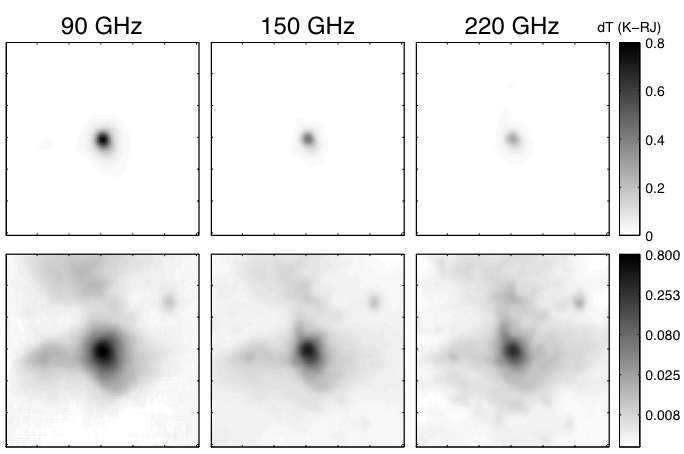}}}
\caption{SPT images of the galactic HII region RCW38.  The top row shows a
linear scale and the bottom row shows a logarithmic scale.  The extent
of each image is $30^{'}$ x $30^{'}$.  Each image is centered on R.A.
$8^{h}$ $59^{m}$ $06^{s}$, Decl. $-47^{\circ}$ $30^{'}$ $38^{''}$.
Increasing R.A. is to the left and increasing Decl. is up.}
\label{rcw38_image}
\end{figure}

%\bibliography{../../BIBTEX/spt.bib}

\end{document}